\documentclass{jpp}

\usepackage{graphicx}
\usepackage{natbib}

\ifCUPmtlplainloaded \else
  \checkfont{eurm10}
  \iffontfound
    \IfFileExists{upmath.sty}
      {\typeout{^^JFound AMS Euler Roman fonts on the system,
                   using the 'upmath' package.^^J}%
       \usepackage{upmath}}
      {\typeout{^^JFound AMS Euler Roman fonts on the system, but you
                   dont seem to have the}%
       \typeout{'upmath' package installed. JPP.cls can take advantage
                 of these fonts, if you use 'upmath' package.^^J}%
      }
  \else
  \fi
\fi

\ifCUPmtlplainloaded \else
  \checkfont{msam10}
  \iffontfound
    \IfFileExists{amssymb.sty}
      {\typeout{^^JFound AMS Symbol fonts on the system, using the
                'amssymb' package.^^J}%
       \usepackage{amssymb}%

      }{}
  \fi
\fi

\ifCUPmtlplainloaded \else
  \IfFileExists{amsbsy.sty}
    {\typeout{^^JFound the 'amsbsy' package on the system, using it.^^J}%
     \usepackage{amsbsy}}
    {\providecommand\boldsymbol[1]{\mbox{\boldmath $##1$}}}
\fi

\newsavebox{\astrutbox}
\sbox{\astrutbox}{\rule[-5pt]{0pt}{20pt}}

\title[Phase Diagrams of Forced Magnetic Reconnection in Taylor's Model]{Phase Diagrams of Forced Magnetic Reconnection in Taylor's Model}

\author[L. Comisso, D. Grasso and F.L. Waelbroeck]%
{L.\ns C\ls O\ls M\ls I\ls S\ls S\ls O$^1$%
  \thanks{Contributed talk at the International Workshop ``Complex Plasma Phenomena in the Laboratory and in the Universe''. Email address for correspondence: luca.comisso@polito.it},\ns
D.\ns G\ls R\ls A\ls S\ls S\ls O$^1$
\and F.\ns L.\ns W\ls A\ls E\ls L\ls B\ls R\ls O\ls E\ls C\ls K$^2$}

\affiliation{$^1$Dipartimento Energia, Politecnico di Torino, Corso Duca degli Abruzzi 24, Torino, 10129, Italy, and Istituto dei Sistemi Complessi - CNR, Via dei Taurini 19, Roma, 00185, Italy\\[\affilskip]
$^2$Institute for Fusion Studies, The University of Texas at Austin, Austin, TX 78712-1203, USA}

\begin{document}

\maketitle

\begin{abstract}

Recent progress in the understanding of how externally driven magnetic reconnection evolves is organized in terms of parameter space diagrams. These diagrams are constructed using four pivotal dimensionless parameters: the Lundquist number $S$, the magnetic Prandtl number $P_m$, the amplitude of the boundary perturbation $\hat \Psi_0$, and the perturbation wave number $\hat k$. This new representation highlights the parameters regions of a given system in which the magnetic reconnection process is expected to be distinguished by a specific evolution. Contrary to previously proposed phase diagrams, the diagrams introduced here take into account the dynamical evolution of the reconnection process and are able to predict slow or fast reconnection regimes for the same values of $S$ and $P_m$, depending on the parameters that characterize the external drive, never considered so far. These features are important to understand the onset and evolution of magnetic reconnection in diverse physical systems.
\end{abstract}

\section{Introduction}

Magnetic reconnection is a process whereby the magnetic field line connectivity \citep[]{Newcomb58, Peg2012, AC_2015} is modified due to the presence of a localized diffusion region. This gives rise to a change in magnetic field line topology and a release of magnetic energy into kinetic and thermal energy. Reconnection of magnetic field lines is ubiquitous in laboratory, space and astrophysical plasmas, where it is believed to play a key role in many of the most striking and energetic phenomena. The most notable examples of such phenomena include sawtooth crashes \citep[]{Yamada94, Nicolas2012} and major disruptions in tokamak experiments \citep[]{Waddell78, Boozer2012}, solar and stellar flares \citep[]{Masuda1994, Su2013}, coronal mass ejections \citep[]{LF_2000, Murphy2012}, magnetospheric substorms \citep[]{Oier2001, Eastwood2007}, coronal heating \citep[]{Priest98, Cassak2008}, and high-energy emissions in pulsar wind nebulae, gamma-ray bursts and jets from active galactic nuclei \citep[]{Kagan2015, Sironi2015, Guo2015}. An exhaustive understanding of how magnetic reconnection proceeds in various regimes is therefore essential to shed light on these phenomena.

In recent years, for the purpose of organizing the current knowledge of the reconnection dynamics that is expected in a system with given plasma parameters, a particular form of phase diagrams have been developed \citep[]{JD_2011, HBS_2011, DR_2012, HB_2013, CD_2013, KA_2013}. These diagrams classify what ``phase'' of magnetic reconnection should occur in a particular system, which is identified by two dimensionless plasma parameters, the Lundquist number
\begin{equation}
S_{L_s} \equiv \frac{{{L_s}{v_{A,u}}}}{{{D_\eta }}} \, , \label{def_S}
\end{equation}
and the macroscopic system size
\begin{equation}
\Lambda  \equiv \frac{L_s}{{{l_k}}} \, . \label{def_lambda}
\end{equation}
Here, $L_s$ indicates the system size in the direction of the reconnecting current sheet, $v_{A,u}$ is the Alfv\'en speed based on the reconnecting component of the magnetic field upstream of the diffusion region, $D_\eta = \eta c^2 /4\pi$ is the magnetic diffusivity, and $l_k$ is the relevant kinetic length scale. This length scale corresponds to \citep[see, e.g.,][]{SimaChac2008, Comisso2013}
\begin{equation}
  l_k = \left\{
    \begin{array}{ll}
      d_i=c/\omega_{pi}      & \mbox{for antiparallel reconnection}, \\[2pt]
      \rho_\tau=c_s/\omega_{ci}    & \mbox{for guide-field reconnection}.
    \end{array} \right.
\end{equation}
Of course, $\omega_{pi}$ is the ion plasma frequency, $\omega_{ci}$ is the ion cyclotron frequency, and $c_s$ is the sound speed based on both the electron and ion temperatures.

All the proposed phase diagrams \citep[]{JD_2011, HBS_2011, DR_2012, HB_2013, CD_2013, KA_2013} exhibit a strong similarity and only a few minor differences. They are useful to summarize some of the current knowledge of the magnetic reconnection dynamics, but they lack fundamental aspects that can greatly affect the reconnection process (some caveats in the use of these diagrams have been discussed by \citet[]{CD_2013}). For example, they do not take into account the dependence of the reconnection process on the external drive or on the magnetic free energy available in the system. An attempt to include these effects has been discussed by \citet[]{JD_2011}, who proposed to incorporate them by adjusting the definition of the Lundquist number, Eq. (\ref{def_S}), but this solution should be viewed only as a rough way to circumnavigate the problem. A further issue is that these diagrams do not consider the evolution of the reconnection process and predict reconnection rates wich are always fast (the estimated reconnection inflow is always a significant fraction of $v_{A,u}$). This, however, in not what is commonly observed in laboratory, space, and astrophysical plasmas, where magnetic reconnection exhibits disparate time scales and is often characterized by an impulsive behaviour, i.e., a sudden increase in the time derivative of the reconnection rate \citep[see, e.g.,][]{Bhatta_2004, Yam_2011}.

Here we propose a different point of view in which we include explicitly the effects of the external drive and the plasma viscosity (neglected in all previous diagrams) on the magnetic reconnection process by considering a four-dimensional parameter space. Then, in this four-dimensional diagram we identify specific domains of parameters where the reconnection process exhibits distinct dynamical evolutions. In other words, in each of these domains the reconnection process goes through diverse phases characterized by different reconnection rates. This analysis leads us to evaluate in greater detail the dynamical evolution of a forced magnetic reconnection process, while collisionless effects have not been taken into account in the present work. 
We introduce the considered model of forced magnetic reconnection in Sec. \ref{sec:taylor_model}, whereas Sec. \ref{sec:conditions} is devoted to the presentation of the possible evolutions of the system and the conditions under which these different evolutions occur. In Sec. \ref{sec:phase_diagrams} we construct the parameter space diagrams that show which reconnection evolution is expected in a system with given characteristic parameters. Finally, open issues are discussed in Sec. \ref{sec:discussion}.

\section{Forced magnetic reconnection in Taylor's model} \label{sec:taylor_model}

\begin{figure}
\begin{center}
\includegraphics[bb = 0 0 198 155, height=6.0cm]{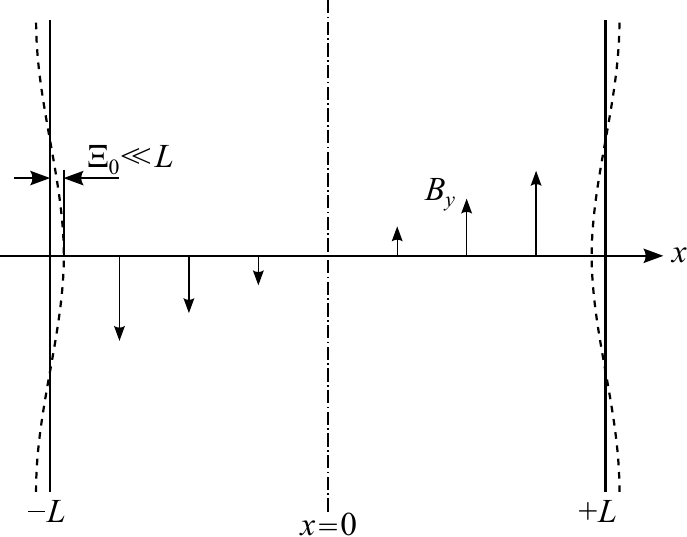}
\end{center}
\caption{Geometry of the Taylor model. The equilibrium magnetic field component $B_y$ is sheared in the $x$ direction, being null at $x=0$. The plasma is bounded by perfectly conducting walls at $x=\pm L$, while it is periodic in the $y$ direction. Magnetic reconnection is driven at $x=0$ by the perturbation $\Xi_0 \cos (ky)$ at the perfectly conducting walls.}
\label{fig1}
\end{figure}

Magnetic reconnection in a given system is conventionally categorized as spontaneous or forced. Spontaneous magnetic reconnection refers to the case in which the reconnection arises by some internal instability of the system or loss of equilibrium, with the most typical example being the tearing mode. Forced magnetic reconnection instead refers to the cases in which the reconnection is driven by some externally imposed flow or magnetic perturbation. In this case, one of the most important paradigms is the so-called ``Taylor problem'', which consists in the study of the evolution of the magnetic reconnection process in a tearing-stable slab plasma equilibrium which is subject to a small amplitude boundary perturbation. This situation is depicted in Fig. \ref{fig1}, where the shared equilibrium magnetic field has the form
\begin{equation}
\boldsymbol{B} = {B_z}{\boldsymbol{e}_z} + (x/L){B_0}{\boldsymbol{e}_y} \, , \label{e_B}
\end{equation}
with $B_z$, $B_0$ and $L$ as constants, and the perfectly conducting walls which bound the plasma are located at $x=\pm L$. Magnetic reconnection is driven at the resonant surface $x=0$ by a deformation of the conducting walls such that 
\begin{equation}
x_w \to \pm L \mp \Xi_0 \cos (ky) \, ,
\end{equation}
where $k=2\pi/ L_y$ is the perturbation wave number and $\Xi_0$ is a small ($\ll L$) displacement amplitude. 
The boundary perturbation is assumed to be set up in a time scale that is long compared to the Alfv\'en time $\tau_A = L/ v_A$, with $v_A=B_0/\sqrt {4\pi \rho}$, but short compared to any characteristic reconnection time scale. Hence, the plasma can be considered in magnetostatic equilibrium everywhere except near the resonant surface at $x=0$.

The first and probably most important contribution to unveiling the behaviour of forced magnetic reconnection in Taylor's model is due to \citet[]{HK_1985}, who showed that very small amplitude boundary perturbations cause an initial linear phase in which a current sheet builds up at the resonant surface, and successive phases in which the reconnection process evolves according to a linear resistive regime and a nonlinear Rutherford regime \citep[]{Rutherford_1973}. The scenario discussed by \citet[]{HK_1985}, which is characterized by a very slow evolution of the reconnection process, was complemented some years later by \citet[]{WB_1992}, who showed that larger perturbations may foster reconnection to proceed through the nonlinear regime according to a Sweet-Parker-like evolution \citep[]{Waelbroeck_1989}, which only on the long time scale of resistive diffusion gives way to a Rutherford evolution. The scenario outlined by \citet[]{WB_1992} is characterized by a reconnection evolution faster than that presented by \citet[]{HK_1985}, but it could still be slow for very small values of plasma resistivity, since in both the Sweet-Parker-like \citep[]{Waelbroeck_1989} and Rutherford \citep[]{Rutherford_1973} regimes, the reconnection rate is strongly dependent on the resistivity, which is known to be extremely small in many laboratory fusion plasmas and space/astrophysical plasmas. However, recent works \citep{Comisso2014, Comisso2015} have shown that relatively large boundary perturbations lead to a different reconnection evolution in plasmas with small resistivity and viscosity. In these cases, after a linear inertial phase and an initial nonlinear regime characterized by a gradually evolving current sheet, the reconnection suddenly enters into a fast reconnection regime distinguished by the disruption of the current sheet due to the development of secondary magnetic islands (usually called plasmoids \citep{Bis_2000,Lou_2007}).

In addition to the works discussed above, which adopt a magnetohydrodynamic (MHD) description of the plasma, we emphasize that many other efforts have been devoted to investigate the Tayor problem assuming MHD, two-fluid and kinetic descriptions \citep[see][]{WangBhatta_1992,Ma_1996,Rem_1998,Vekstein_1998,
Avinash_1998,Vekstein_1999,Valori_2000,Fitz_2003,FBML_2003,Fitz_2004a,Fitz_2004b,
Cole_2004,Bian_2005,Birn_2005,Vekstein_2006,Birn_2007,Fitz_2008,HossVek2008,Gordo_2010a,
Gordo_2010b,LazzCom2011,Hosseinpour2013,Dewar_2013}. 
Indeed, the Taylor problem has important applications besides being interesting from the point of view of basic physics. For instance, in laboratory fusion plasmas the Taylor model represents a convenient  way to study magnetic reconnection processes driven by resonant magnetic perturbations, while in astrophysical plasmas this model can be adopted to study magnetic reconnection forced by the motions of photospheric flux tubes.

\section{Evolution of the reconnection process in Taylor's model} \label{sec:conditions}

In this section we review the present understanding of the forced magnetic reconnection dynamics in Taylor's model focusing on a visco-resistive plasma with $P_m$ greater than $1$. As shown by \citet[]{HK_1985}, this dynamics always starts with a linear inertial phase in which a current sheet builds up at the resonant surface and shrinks inversely in time. Concurrently, the current density at the $X$-point increases linearly in time. The reconnection rate during this phase can be evaluated by recalling that the current density is proportional to the out-of-plane electric field at the $X$-point,  which is equal to
\begin{equation}  \label{ReconnRate_IN}
{\left. {{\partial _t}\psi } \right|_X} = \frac{2}{\pi} {\Delta '_s} k L^2 {B_0} {\Xi_0} \frac{t}{\tau_A \tau_\eta}
\end{equation}
for $t \ll \tau_\nu^{1/3} {({\tau _A}/kL)^{2/3}}$ \citep[]{Fitz_2003, Comisso2015}. 
Here, $\psi$ stands for the magnetic flux function of the perturbed magnetic field in the reconnection plane ($\delta \boldsymbol{B}_\bot = \nabla \psi \times {\boldsymbol{e}_z} $), $\tau_\nu = L^2/\nu$ and $\tau_\eta = L^2/ D_\eta$ indicate the characteristic time for viscous and resistive diffusion, respectively, while $\Delta '_s = 2k/\sinh (kL)$ parametrizes the contribution of the external source perturbation to the gradient discontinuity of the magnetic flux function at the resonant surface. It is important to point out that this phase is characterized by a non-constant-$\psi$ behaviour of the magnetic flux function across the island. Depending on whether or not this property persists until the beginning of the nonlinear regime, different scenarios may occur.

\subsection{Hahm-Kulsrud scenario} \label{secHK}

If the boundary perturbation is such that \citep[]{Fitz_2003, Comisso2015}
\begin{equation}
\Psi_0 = {B_0}{\Xi_0} \ll {\left( {{\tau _\nu }{\tau _\eta }} \right)^{ - 1/6}}{\left( {\frac{{{\tau _A}}}{{kL}}} \right)^{1/3}}\frac{{{B_0}}}{{\Delta '_s }}  \equiv \Psi_W \, ,
\end{equation}
after the inertial phase the reconnection process evolves trough a visco-resistive phase, which is a linear regime characterized by a constant-$\psi$ behaviour, i.e., the perturbed magnetic flux function can be treated as a constant in $x$ over the width of the reconnection layer. During this phase the reconnection rate is given by 
\begin{equation}  \label{ReconnRate_VR}
{\left. {{\partial _t}\psi } \right|_X} = {B_0}{\Xi_0}\frac{{{\Delta '_s}L}}{{{\tau _*}}}{e^{{\Delta '_0}Lt/{\tau _*}}} \, ,
\end{equation}
where $\Delta '_0 = 2k/\tanh (kL)$ is the standard tearing stability parameter and $\tau _*$ is a characteristic time defined as \citep[]{Fitz_2003, Comisso2015}
\begin{equation} 
{\tau _*} \equiv \pi {6^{2/3}}\frac{{\Gamma \left( {\frac{5}{6}} \right)}}{{\Gamma \left( {\frac{1}{6}} \right)}}\frac{{\tau _\eta ^{5/6}}}{{\tau _\nu ^{1/6}}}{\left( {\frac{{{\tau _A}}}{{kL}}} \right)^{1/3}} \, ,
\end{equation}
with $\Gamma$ indicating the Gamma function. Eq. (\ref{ReconnRate_VR}) is valid for $t \gg \tau_\nu^{-1/3}\tau_\eta^{2/3}{({\tau_A}/kL)^{2/3}}$ \citep[]{Fitz_2003, Comisso2015} and a magnetic island width much smaller than the linear layer width, i.e., $w \ll \delta_{\nu \eta} \sim {({\tau _\nu }{\tau _\eta })^{-1/6}}{({\tau _A}/kL)^{1/3}}L$ \citep[]{Porcelli_1987, Fitz_NF1993}. 
If the perturbation is sufficient to drive the magnetic island into the nonlinear regime ($w \gtrsim \delta_{\nu \eta}$), the visco-resistive phase ends up into a Rutherford evolution, whose island width growth is governed by the Rutherford equation
\begin{equation}  \label{Rutherford_Eq}
{\cal I} \frac{\tau_\eta}{L^2} \frac{dw}{dt} = \Delta '_0 + \Delta '_s  \frac{{{\Psi _0}}}{{{{\left. \psi  \right|}_X}}} \, ,
\end{equation}
where ${\cal I}=0.823$ \citep[]{Rutherford_1973, Fitz_NF1993}. This is a very slow reconnection evolution in which the reconnection rate can be evaluated analytically in the two limits \citep[]{HK_1985, Comisso2015} 
\begin{equation}  \label{ReconnRate_RU1}
{\left. {{\partial _t}\psi } \right|_X} = \frac{{2{\Delta '_s}{\Psi _0}}}{{( - \Delta '_0){\tau_{NL}}}}{\left( {\frac{{3t}}{{{\tau_{NL}}}}} \right)^{-1/3}} \quad \mbox{for} \quad  t \ll \tau_{NL} \, ,
\end{equation}
\begin{equation}  \label{ReconnRate_RU2}
{\left. {{\partial _t}\psi } \right|_X} = \frac{{2{\Delta '_s}{\Psi _0}}} {{( - \Delta '_0){\tau_{NL}}}} \tanh \left( {\frac{t}{{{\tau_{NL}}}}} \right) \cosh ^{-2} \left( {\frac{t}{{{\tau_{NL}}}}} \right)  \quad \mbox{for} \quad  t \gg \tau_{NL} \, ,
\end{equation}
where 
\begin{equation}  \label{}
\tau_{NL} = \frac{{4 {\cal I}}}{{( -\Delta '_0)L}} {\left( {\frac{{{\Delta '_s}}}{{( - {\Delta '_0})}}\frac{{{\Xi _0}}}{L}} \right)^{1/2}}{\tau _\eta}  \, .
\end{equation}

\subsection{Wang-Bhattacharjee scenario} \label{secWB}

If the boundary perturbation is such that \citep[]{Fitz_2003, Comisso2015}
\begin{equation}
\Psi_0 \gtrsim \Psi_W  \, ,
\end{equation}
the non-constant-$\psi$ behaviour characteristic of the inertial phase lingers until the nonlinear regime is entered. Therefore, since in this case the magnetic island grows faster than the current can diffuse out of the reconnecting layer, the evolution of the reconnection process is distinguished by a strong current sheet at the resonant surface \citep[]{Waelbroeck_1989}. The reconnecting current sheet turns out to be stable if the boundary perturbation is such that \citep[]{Comisso2015}
\begin{equation}  \label{Plasmoid_Condition_3}
\Psi_0 = {B_0}{\Xi _0} <  C {B_0}L \frac{k}{\Delta '_s} \frac{{{\tau _A}}}{{{\tau _\eta }}}{\left( {1 + \frac{{{\tau _\eta }}}{{{\tau _\nu }}}} \right)^{1/2}} \equiv \Psi_c  \, ,
\end{equation}
where the multiplicative constant $C \sim 2 \epsilon_c^{-2}$ depends on the critical inverse aspect ratio of the reconnecting current sheet (specified later). In this case, the reconnection process follows a Sweet-Parker evolution (modified by plasma viscosity \citep[]{Park_1984}), whose reconnection rate in Taylor's model is \citep[]{Comisso2015}
\begin{equation}  \label{Rec_rate_at_X_4}
{\left. {\partial_t}{\psi} \right|_X} \approx \frac{1}{3}{B_0}L{\left( {{\Delta '_s}{\Xi _0}} \right)^{3/2}}{\left( {\frac{{kL}}{{{\tau _A}{\tau _\eta }}}} \right)^{1/2}}{\left( {1 + \frac{{{\tau _\eta }}}{{{\tau _\nu }}}} \right)^{ - 1/4}} \, .
\end{equation}
Finally, the Sweet-Parker type of evolution gives way to a Rutherford evolution on the time scale of resistive diffusion.

\subsection{Our scenario} \label{our_scenario}

If the boundary perturbation satisfies \citep[]{Fitz_2003, Comisso2015}
\begin{equation}
\Psi_0 \gtrsim {\left( {{\tau _\nu }{\tau _\eta }} \right)^{ - 1/6}}{\left( {\frac{{{\tau _A}}}{{kL}}} \right)^{1/3}}\frac{{{B_0}}}{{\Delta '_s }} \equiv \Psi_W \, 
\end{equation}
and also the condition \citep[]{Comisso2015}
\begin{equation}  \label{Plasmoid_Condition_3}
\Psi_0  >  C {B_0}L \frac{k}{\Delta '_s} \frac{{{\tau _A}}}{{{\tau _\eta }}}{\left( {1 + \frac{{{\tau _\eta }}}{{{\tau _\nu }}}} \right)^{1/2}} \equiv \Psi_c  \, ,
\end{equation}
the reconnection process does not reach a stable Sweet-Parker regime, but a different situation occurs. A gradually thinning current sheet evolves until its aspect ratio reaches the limit that allows the plasmoid instability to develop. The growth of the plasmoids leads to the disruption of the current sheet, and therefore to a dramatic increase of the reconnection rate. The reconnection rate during this plasmoid-dominated phase has been evaluated in a statistical steady state as \citep[]{Comisso2015}
\begin{equation} \label{Rec_rate_plasmoids_at_X_3}
{\partial_t}{\psi_p} \approx \epsilon_c {B_0}L{\left( {{\Delta '_s}{\Xi _0}} \right)^2}\tau _A^{ - 1}{\left( {1 + \frac{{{\tau _\eta }}}{{{\tau _\nu }}}} \right)^{-1/2}} \, ,
\end{equation}
where $\epsilon_c = \delta_c / L_c$ is the critical inverse aspect ratio of the reconnecting current sheet. This quantity, whose value has been found to lie in the range $1/100$ - $1/200$ by means of numerical simulations \citep{BHYR_2009,SLUSC_2009,Cassak_2009,Skender_2010,HB_2010}, represents the threshold below which the reconnecting current sheet becomes unstable to the plasmoid instability \citep{Lou_2007}.

\section{Phase diagrams}\label{sec:phase_diagrams}
 
In this section we illustrate the domain of existence of the three different scenarios described before with the help of appropriated parameter space maps. For the sake of clarity we state again the three type of reconnection evolutions we are referring to: \\

(1) Hahm-Kulsrud scenario \citep[]{HK_1985, Fitz_2003}, \\

(2) Wang-Bhattacharjee scenario \citep[]{WB_1992, Fitz_2003}, \\

(3) Our scenario \citep[]{Comisso2015}. \\

\noindent Since each of these scenarios includes different phases/regimes of reconnection, the concept of ``phase diagrams'' is intended here in a broader sense. Due to this fact, they could also be defined in a more general way as ``scenario diagrams''. This kind of diagrams can be constructed from the conditions summarized in the previous section. Therefore, the possible evolutions of the reconnection process may be organized in a four-dimensional parameter space map with $\hat \Psi_0 = \Psi_0 / {B_0} L$, $\hat k=kL$, $S=L v_{A}/{D_\eta}$, and $P_m = \nu/{D_\eta}$ on the four axes. However, due to the difficulty in the visualization of such a four-dimensional diagram, it is convenient to consider two-dimensional slices for fixed values of two of the four parameters.

\begin{figure} 
\begin{center}
\includegraphics[bb = 2 4 358 360, width=6.6cm]{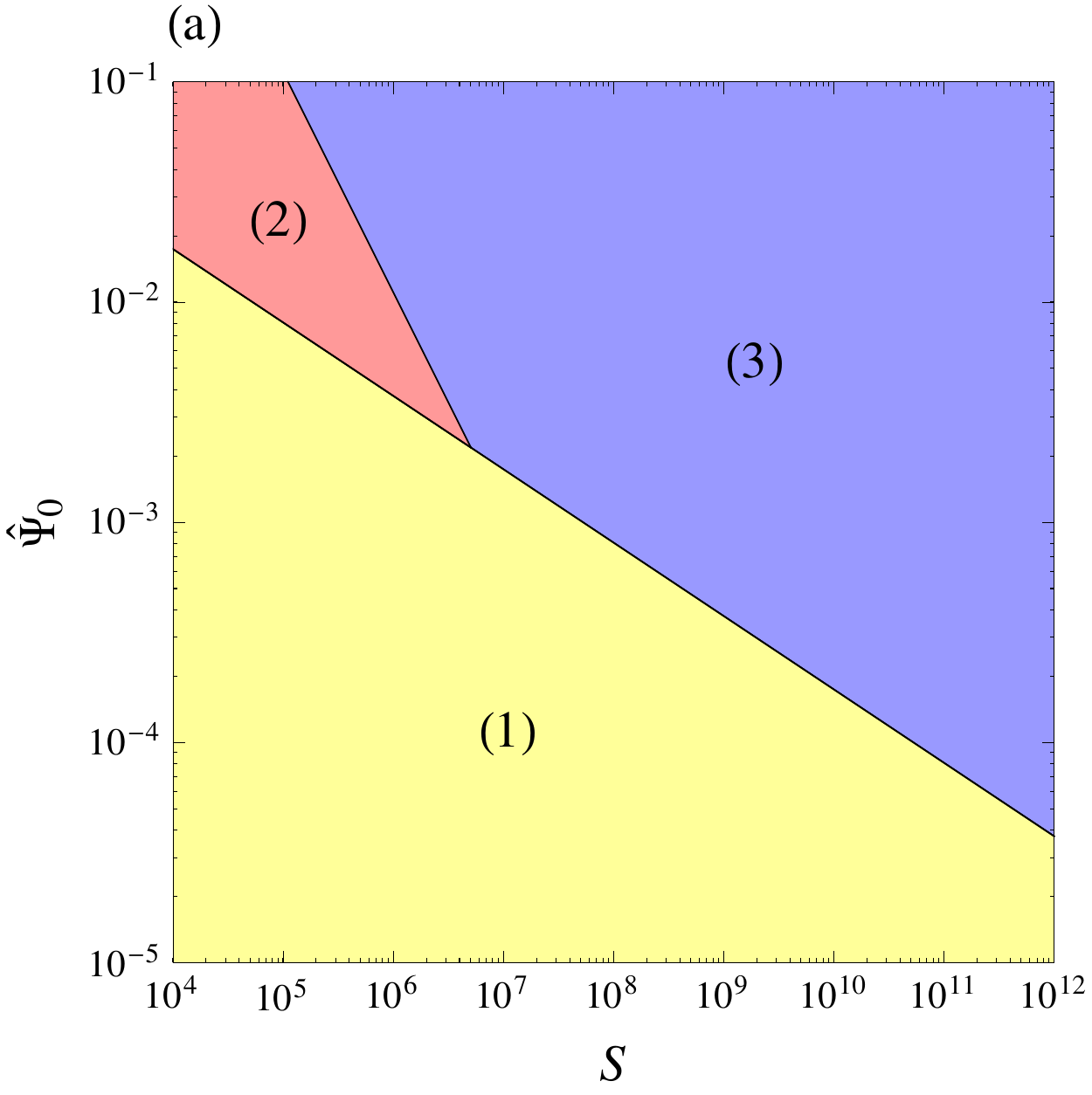}
\includegraphics[bb = 2 4 358 360, width=6.6cm]{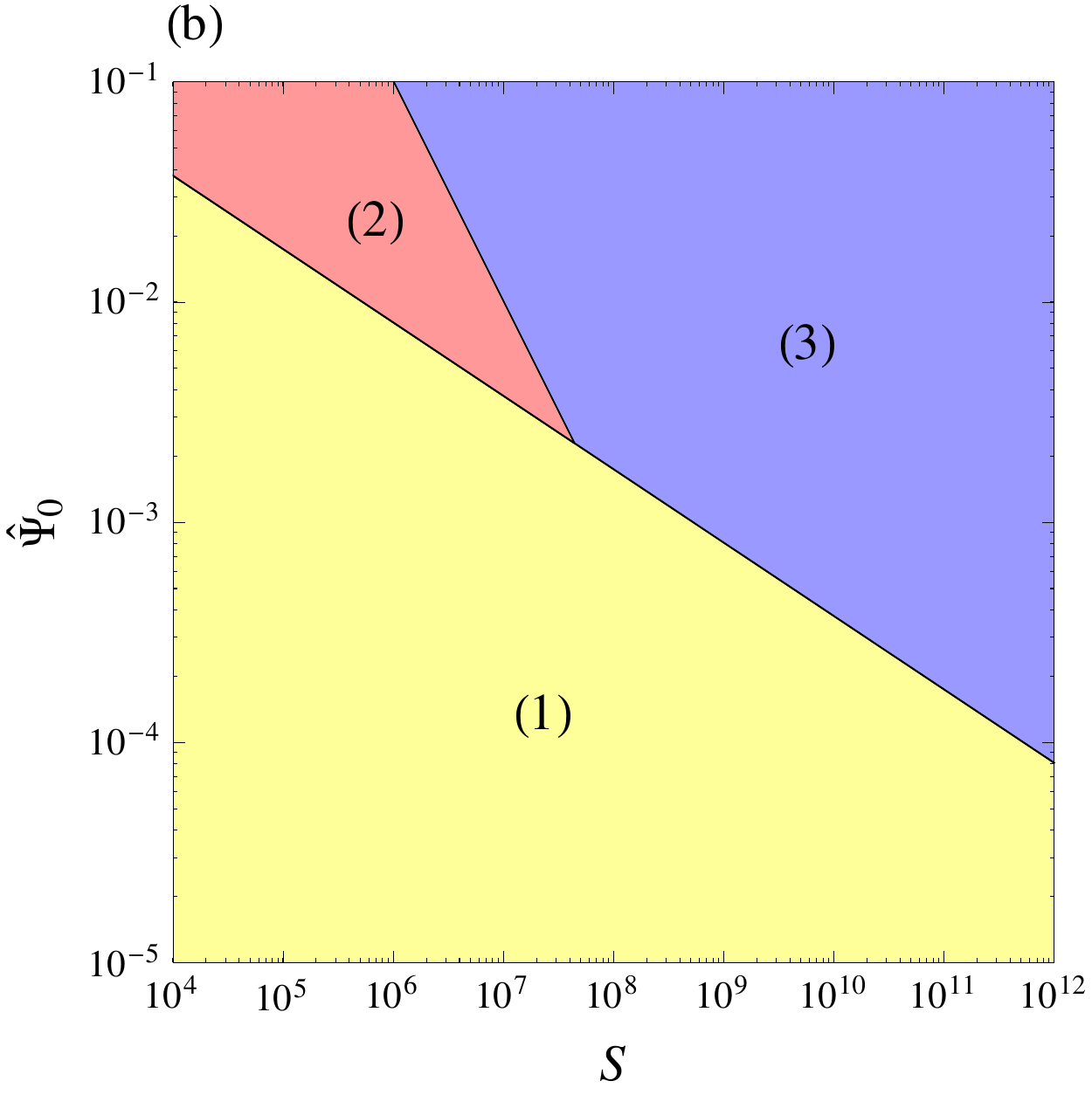}
\includegraphics[bb = 2 4 358 360, width=6.6cm]{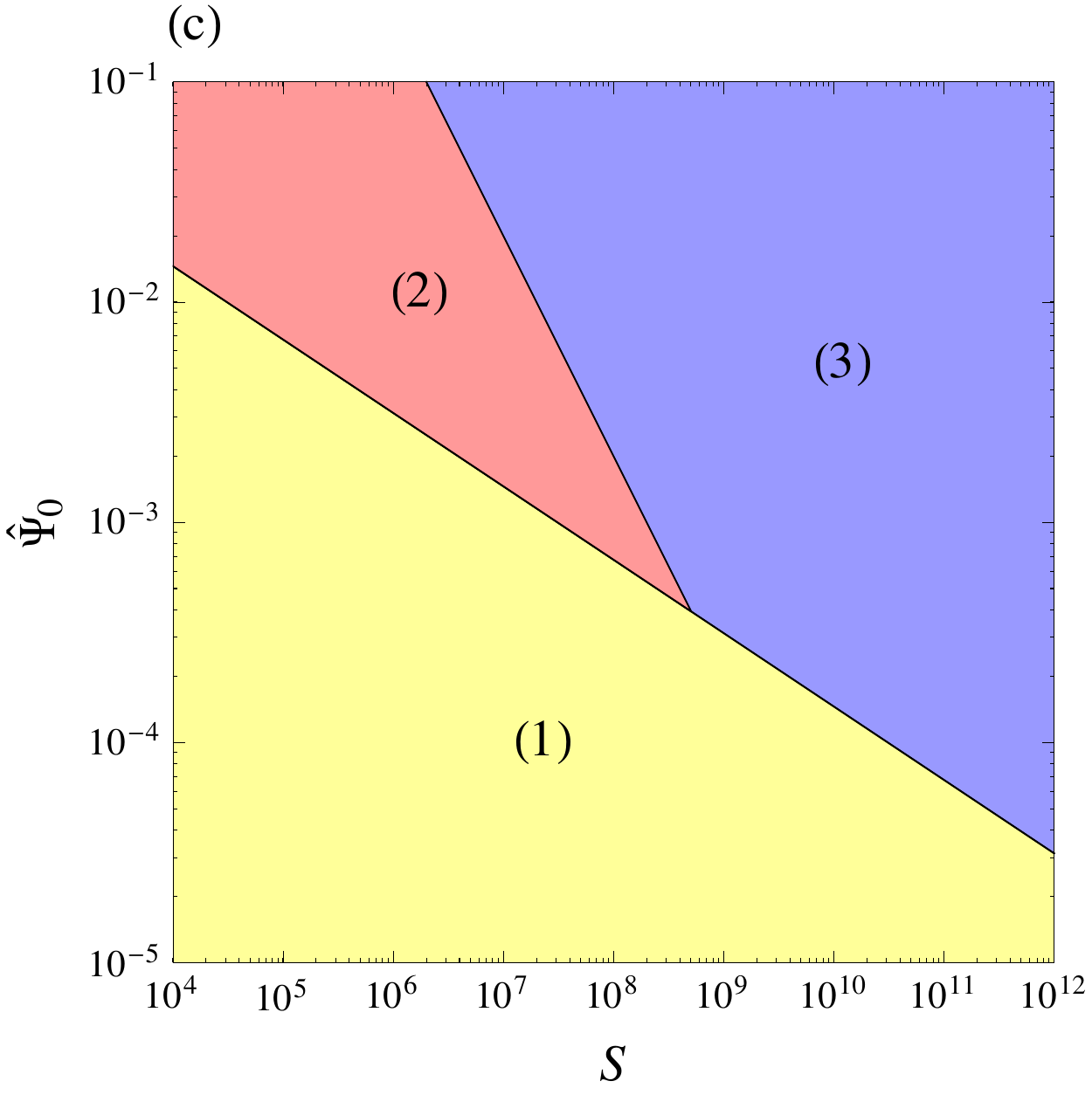}
\includegraphics[bb = 2 4 358 360, width=6.6cm]{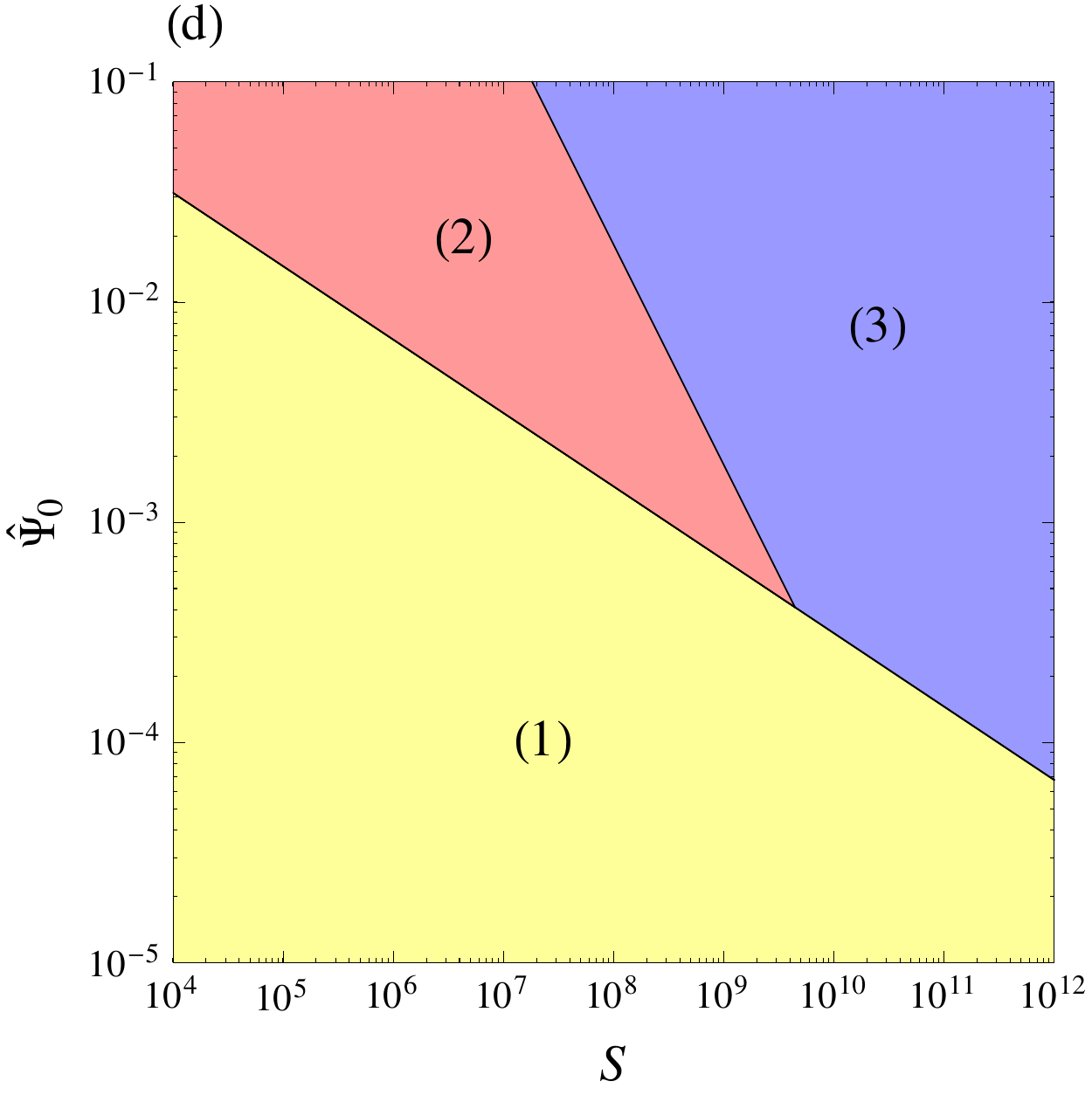}
\end{center}
\caption{Two-dimensional slices of a phase/scenario diagram for forced magnetic reconnection in the magnetohydrodynamical Taylor model. Fixed parameters are (a) $\hat k=1/8$, $P_m=5$, (b) $\hat k=1/8$, $P_m=500$, (c) $\hat k=2$, $P_m=5$, and (d) $\hat k=2$, $P_m=500$. The numerical labels indicate the (1) Hahm-Kulsrud scenario, (2) Wang-Bhattacharjee scenario, and (3) our scenario. The boundaries between the different scenarios are identified by the functions $\hat \Psi_0 = \hat \Psi_W /3$ and $\hat \Psi_0 = \hat \Psi_c$ for $\hat \Psi_c > \hat \Psi_W /3$.}
\label{fig2}
\end{figure}

\begin{figure}
\begin{center}
\includegraphics[bb = 24 1 358 367, height=6.8cm]{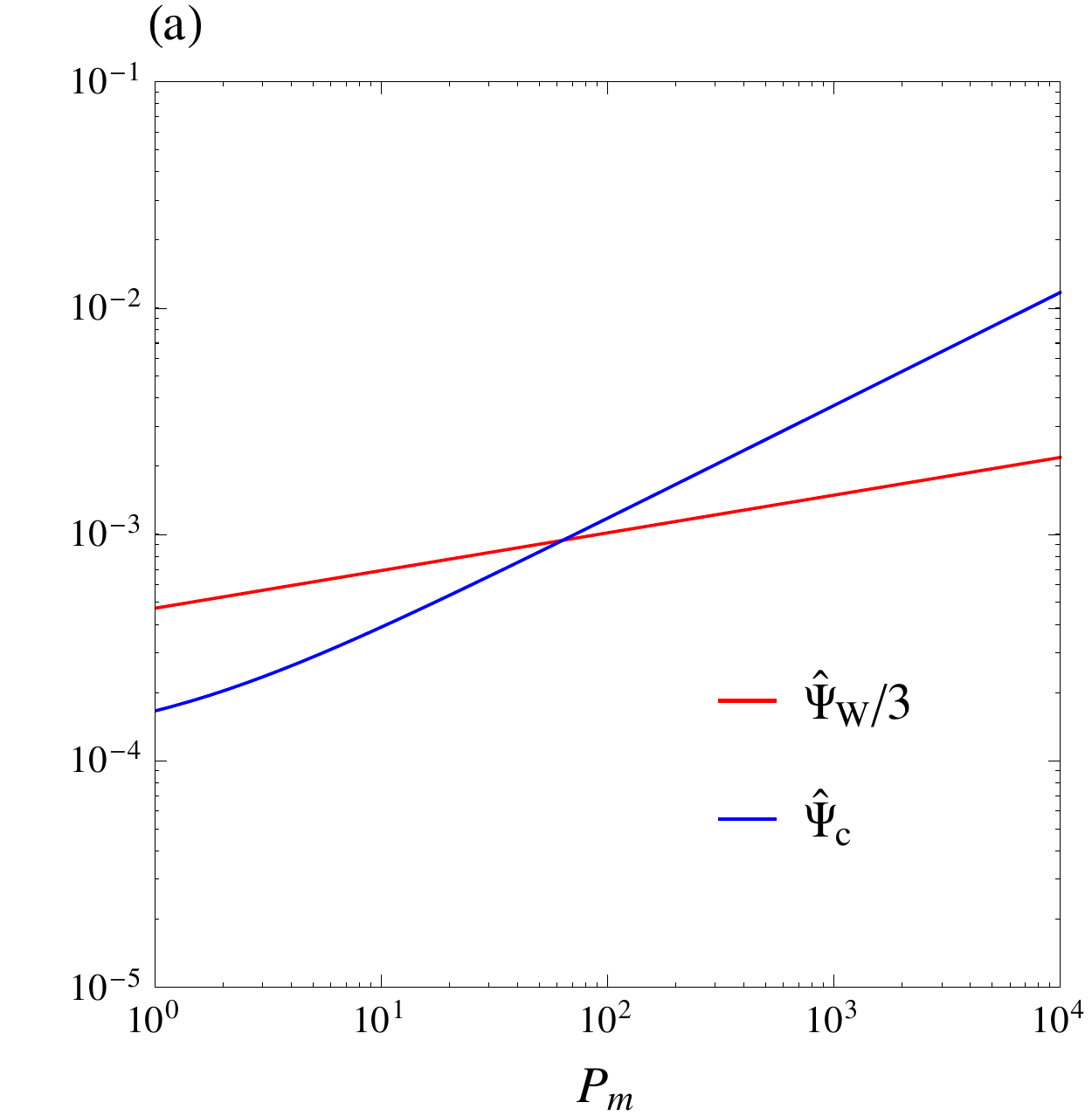}
\includegraphics[bb = 2 2 358 362, height=6.8cm]{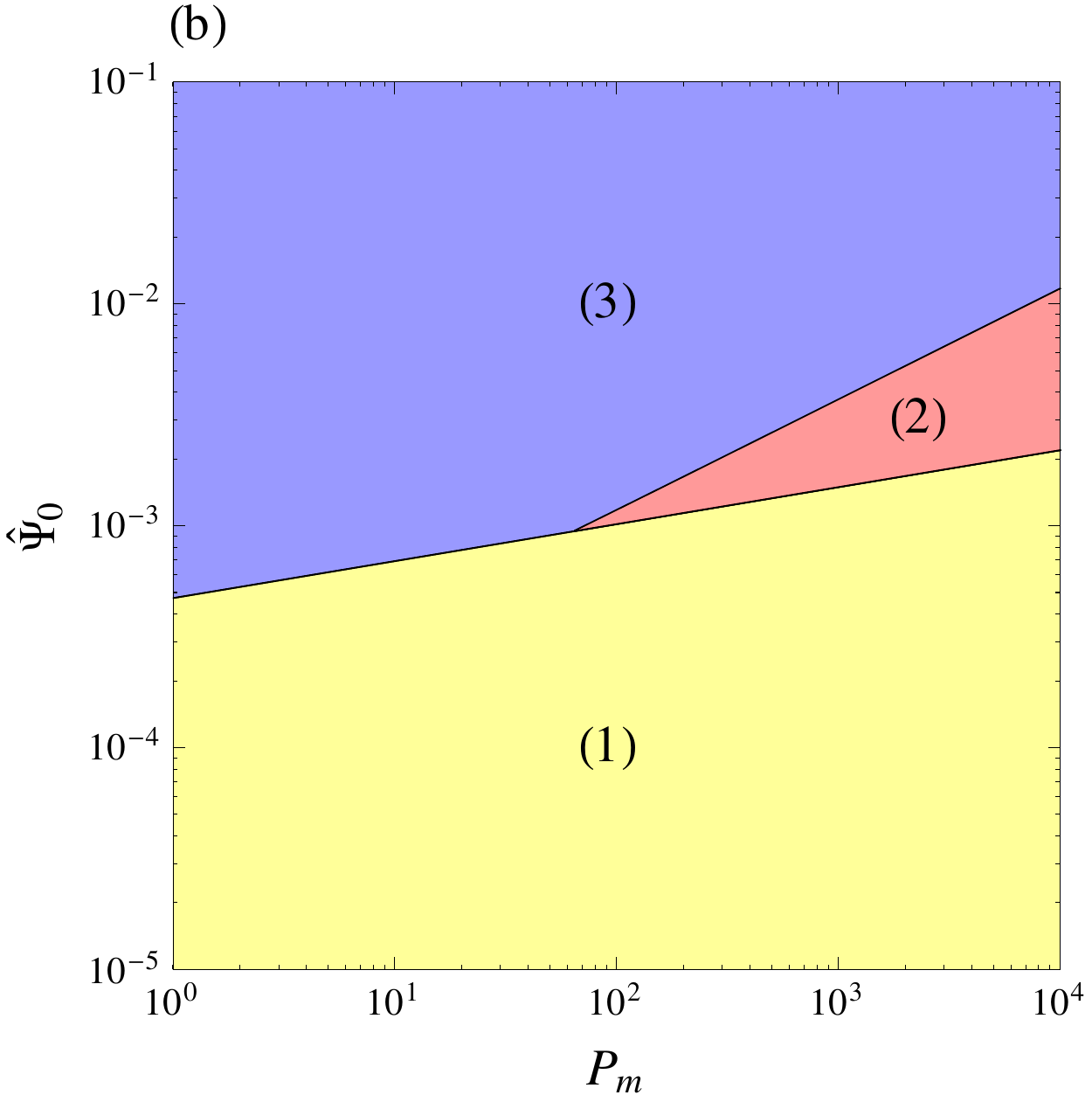}
\end{center}
\caption{(a) Thresholds $\hat \Psi_{W}/3$ (red line) and $\hat \Psi_c$ (blue line) as a function of the magnetic Prandtl number $P_m$ for $S = 10^{8}$, $\hat k = 0.5$ and $C=2(150)^2$. (b) Corresponding two-dimensional slice of the phase/scenario diagram identifying the (1) Hahm-Kulsrud scenario, (2) Wang-Bhattacharjee scenario, and (3) our scenario.}
\label{fig3}
\end{figure}
\begin{figure}
\begin{center}
\includegraphics[bb = 2 5 455 342, height=6.8cm]{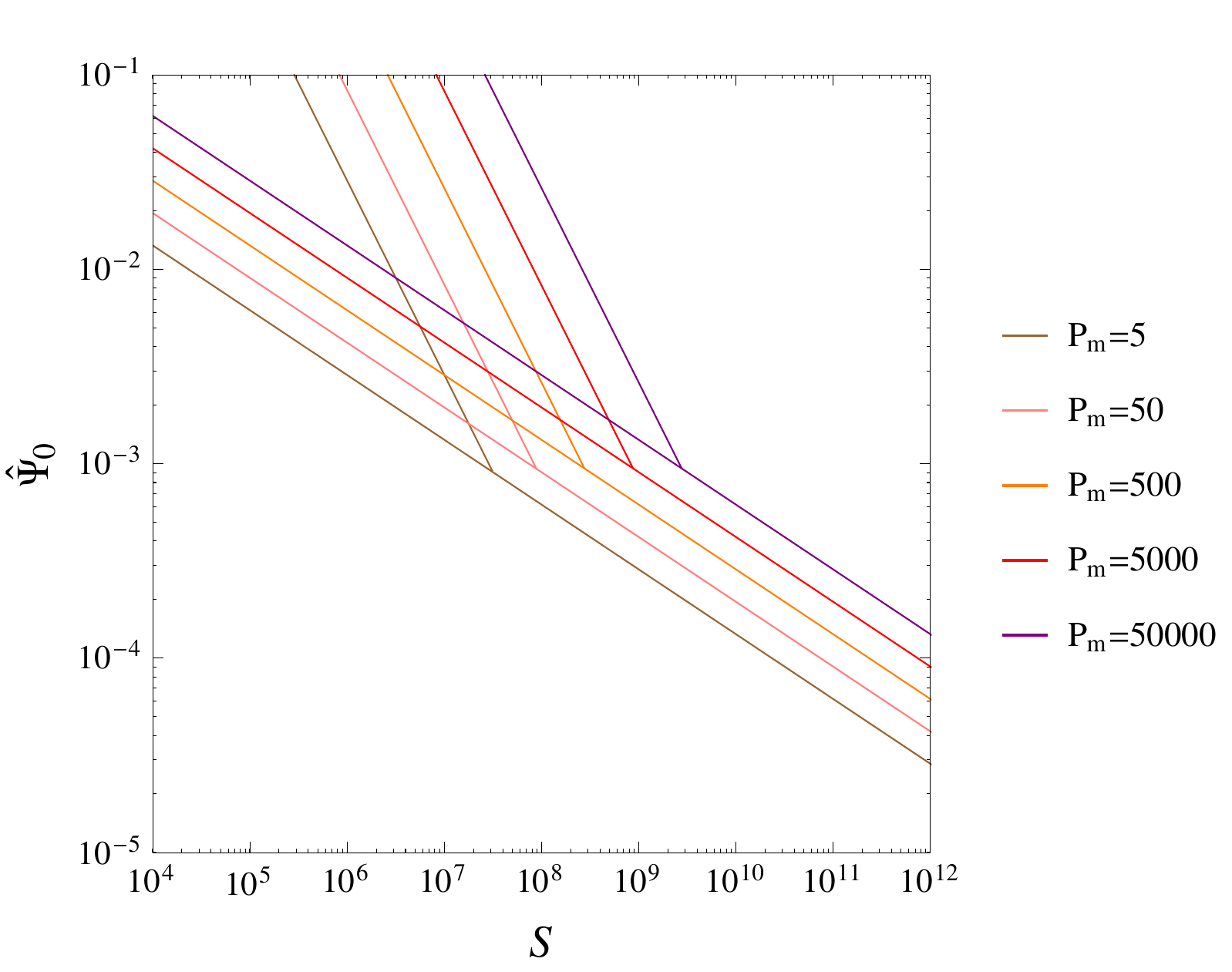}
\end{center}
\caption{Boundaries (identified by the functions $\hat \Psi_0 = \hat \Psi_W /3$ and $\hat \Psi_0 = \hat \Psi_c$ for $\hat \Psi_c > \hat \Psi_W /3$) of the different possible evolutions of the reconnection process for $\hat k = 0.5$ and various values of the magnetic Prandtl number.}
\label{fig4}
\end{figure}

Let us first consider four two-dimensional slices with fixed values of the magnetic Prandtl number and perturbation wave number. Assuming that the Hahm-Kulsrud scenario (which occurs if $\Psi_0 \ll \Psi_W$) may hold until $\Psi_0 = \Psi_W /3$, the corresponding diagrams for (a) $\hat k=1/8$, $P_m=5$, (b) $\hat k=1/8$, $P_m=500$, (c) $\hat k=2$, $P_m=5$, (d) $\hat k=2$, $P_m=500$, are shown in Figs. \ref{fig2}(a) - \ref{fig2}(d). From this plots it is clear that the Wang-Bhattacharjee scenario is limited to a small range of values of the Lundquist number and the source perturbation amplitude. Increasing values of the the magnetic Prandtl number and perturbation wave number extend the domain of existence of this possible type of evolution of the system. However, after a threshold value of the Lundquist number (identified by the intersection of the two black lines representing $\hat \Psi_0 = \hat \Psi_W /3$ and $\hat \Psi_0 = \hat \Psi_c$), the Wang-Bhattacharjee scenario cannot occur because it is not possible to obtain a stable Sweet-Parker-type evolution. In these cases, the Hahm-Kulsrud scenario is facilitated by very small perturbation amplitudes, whereas larger perturbations lead the system to a fast reconnection regime as described in Sec. \ref{our_scenario}. 
Note that while previously proposed phase diagrams always predict fast reconnection \citep[]{JD_2011, HBS_2011, DR_2012, HB_2013, CD_2013, KA_2013}, in clear contrast to what happens in nature, our diagrams show that reconnection proceeds very slowly (region (1)) if the source perturbation is not sufficiently large.

\begin{figure}
\begin{center}
\includegraphics[bb = 24 5 357 368, height=6.8cm]{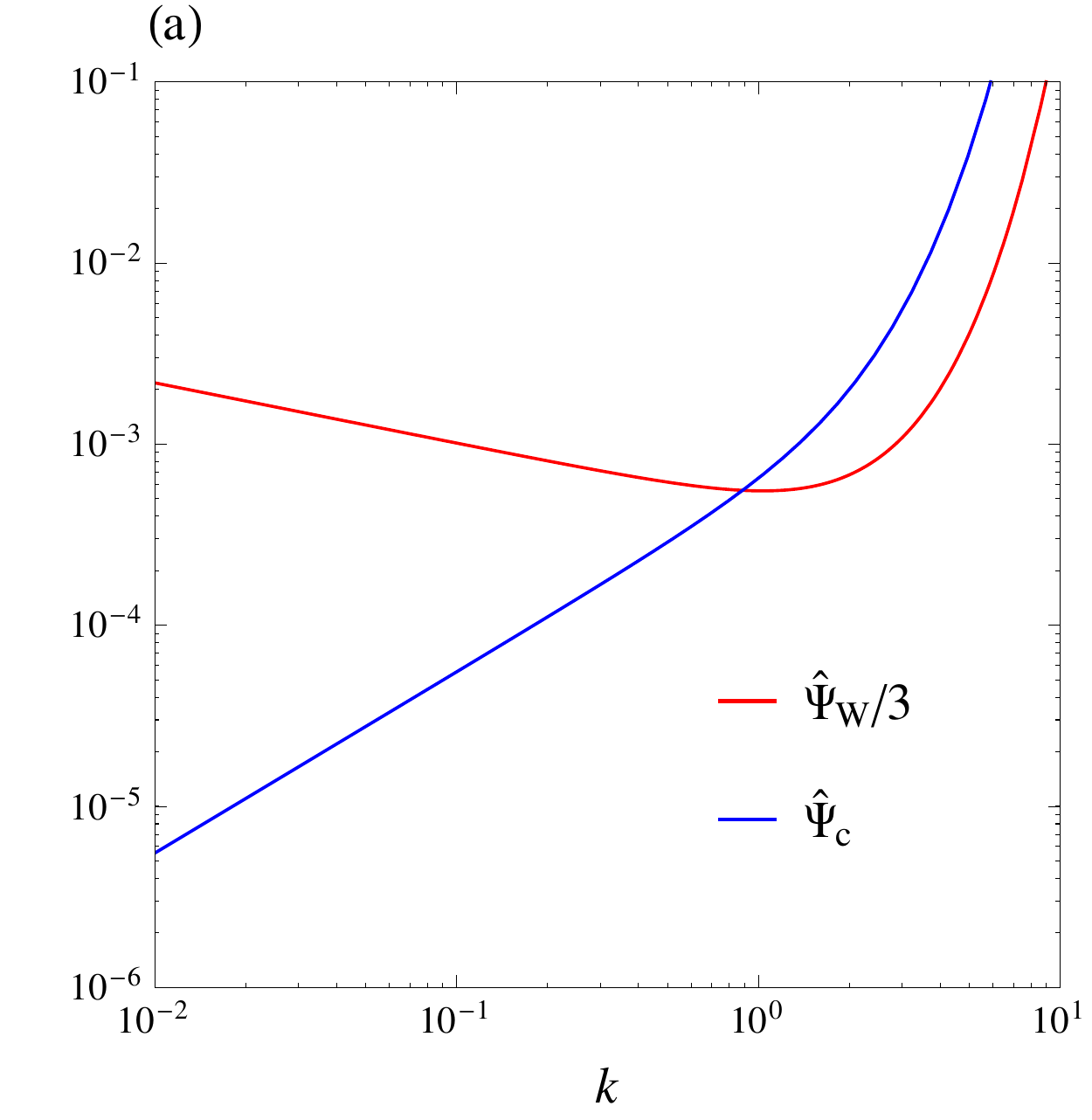}
\includegraphics[bb = 2 5 357 362, height=6.8cm]{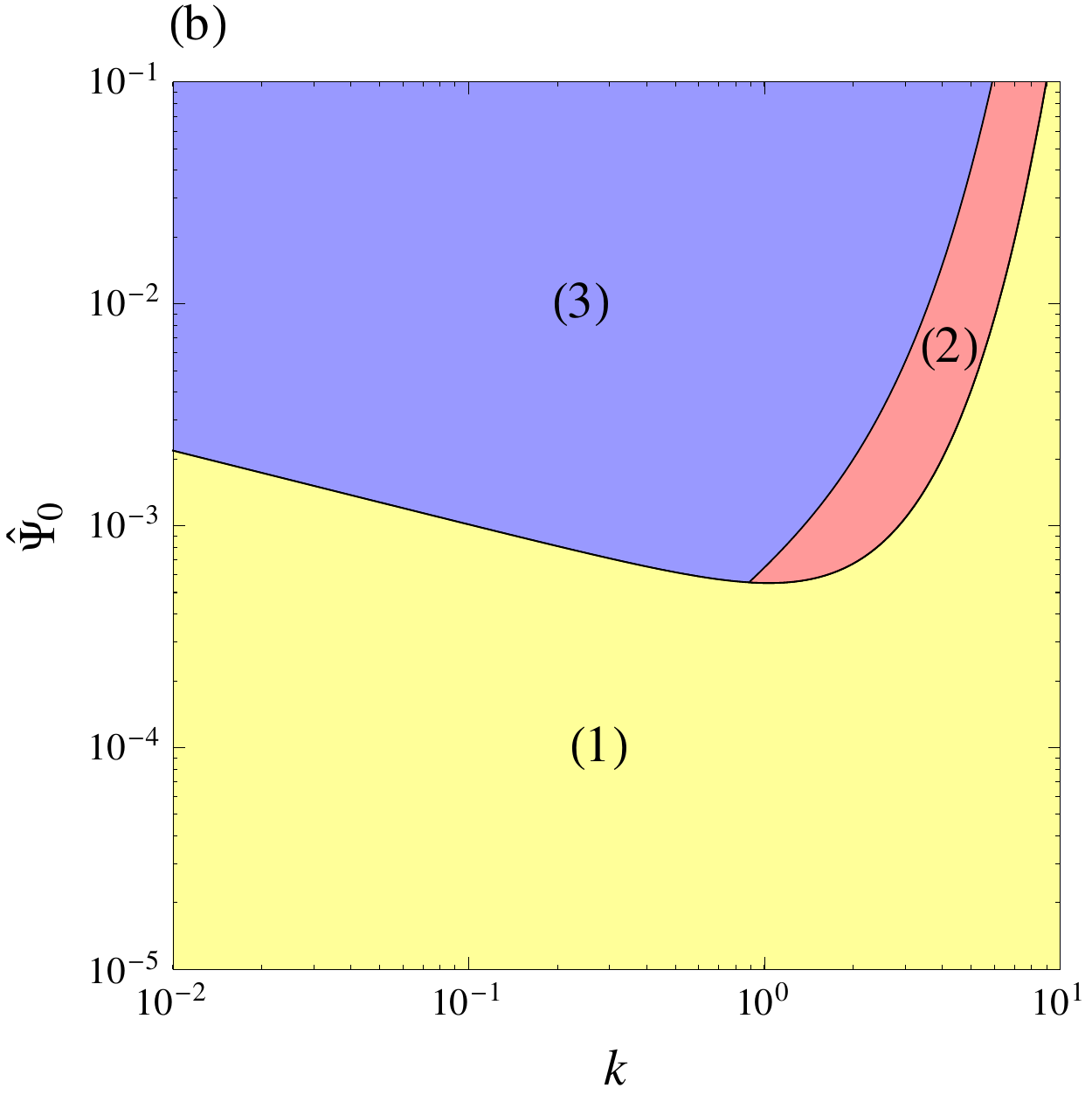}
\end{center}
\caption{(a) Thresholds $\hat \Psi_{W}/3$ (red line) and $\hat \Psi_c$ (blue line) as a function of the perturbation wave number $\hat k$ for $S = 10^{8}$, $P_m = 5$ and $C=2(150)^2$. (b) Corresponding two-dimensional slice of the phase/scenario diagram identifying the (1) Hahm-Kulsrud scenario, (2) Wang-Bhattacharjee scenario, and (3) our scenario.}
\label{fig5}
\end{figure}
\begin{figure}
\begin{center}
\includegraphics[bb = 2 5 443 342, height=6.8cm]{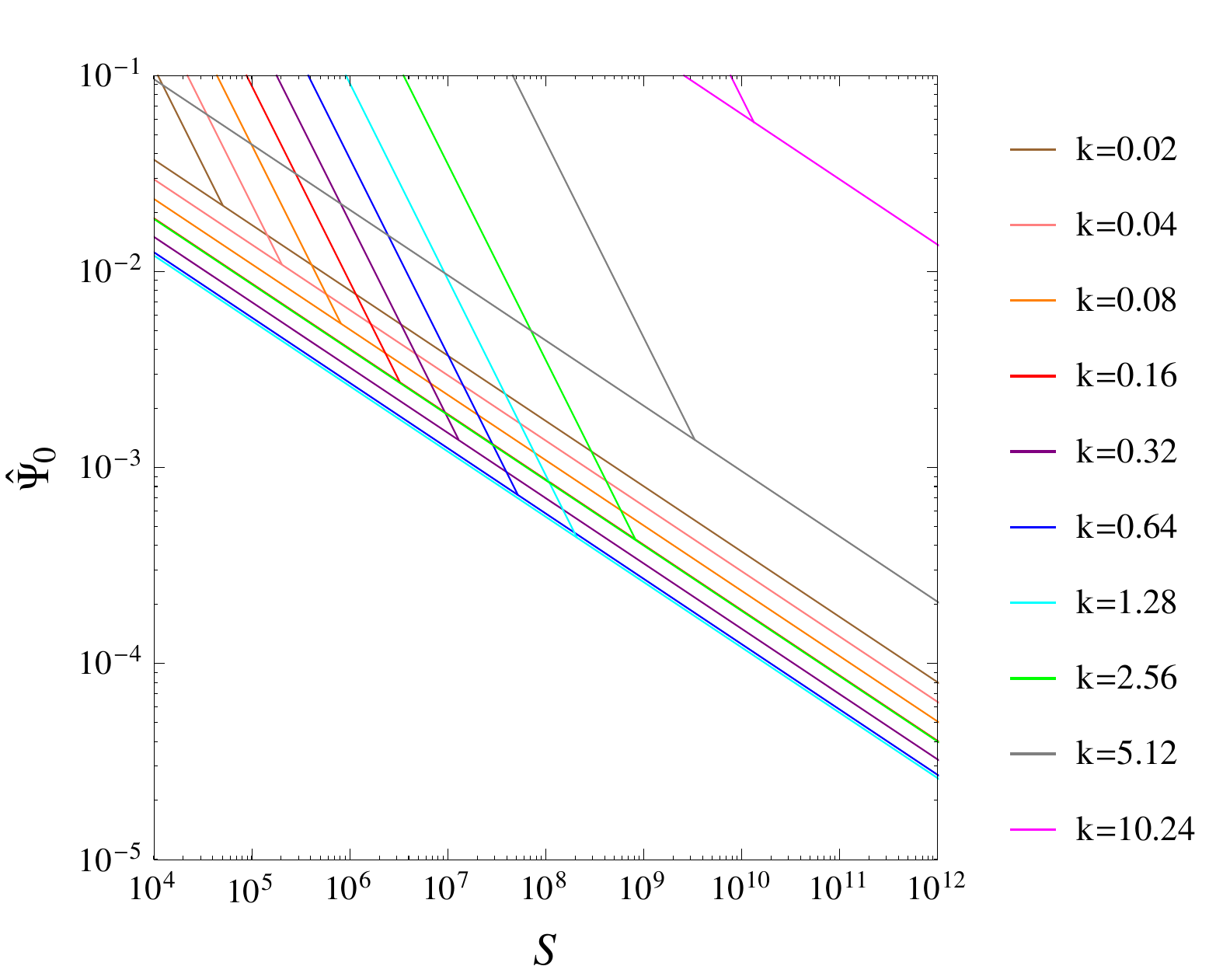}
\end{center}
\caption{Boundaries (identified by the functions $\hat \Psi_0 = \hat \Psi_W /3$ and $\hat \Psi_0 = \hat \Psi_c$ for $\hat \Psi_c > \hat \Psi_W /3$) of the different possible evolutions of the reconnection process for $P_m = 5$ and various values of the perturbation wave number.}
\label{fig6}
\end{figure}

Let us now examine the effect of the plasma viscosity by considering the domain of existence of the different scenarios as a function of the parameters $\hat \Psi_0$ and $P_m$. Fig. \ref{fig3}(a) shows the functions $\hat \Psi_0 = \hat \Psi_W /3$ and $\hat \Psi_0 = \hat \Psi_c$ for $S= 10^{8}$ and $\hat k = 0.5$. For $\hat \Psi_c < \hat \Psi_W /3$ the threshold for the plasmoids formation coincides with that for the nonlinear evolution characterized by a strong reconnecting current sheet. Therefore, for $\hat \Psi_c < \hat \Psi_W /3$ an increase of the amplitude perturbation $\hat \Psi_0$ drives the system directly from scenario (1) to scenario (3). This situation is depicted in Fig. \ref{fig3}(b), where it is clearly shown that the increase of the magnetic Prandtl number has the effect of making possible or extending the domain of existence of scenario (2). 

To clarify the effect of the plasma viscosity we also delineate the boundaries of the diverse evolutions (1)-(3) in a parameter space map $(\hat \Psi_0, S)$ (as in Fig. \ref{fig2}) for fixed $\hat k = 0.5$ but different values of $P_m$. This is shown in Fig. \ref{fig4} for $P_m = 5 - 5 \times {10^5}$. The increase of the magnetic Prandtl number extends the domain of existence of the slow reconnection scenario (1) at the expense of the fast reconnection scenario (3). The area of existence of scenario (2) remains almost unchanged, but shifted towards higher values of the Lundquist number.

We now examine in more detail how the possible evolutions of the forced magnetic reconnection process depend on the wave number of the boundary perturbation. Fig. \ref{fig5}(a) shows the thresholds $\hat \Psi_0 = \hat \Psi_W /3$ and $\hat \Psi_0 = \hat \Psi_c$ as a function of $\hat k$ for fixed values of $S= 10^{8}$ and $P_m = 5$. Below a critical perturbation wave number $\hat k^*$ (corresponding to $\hat k^* \approx 1$ for the fixed parameters used in Fig. \ref{fig5}(a)), every time the non-constant-$\psi$ magnetic island pass into the nonlinear regime, the evolution of the system leads to the plasmoid-dominated phase predicted in scenario (3). The domains of existence of the possible evolutions (1)-(3) are illustrated in Fig. \ref{fig5}(b). The scenario discussed by Wang and Bhattacharjee happens only for a small range of ($\hat \Psi_0, \hat k$) parameters. Not also that scenario (3) is facilitated for $\hat k \lesssim \hat k^*$, while scenario (1) may occur for large amplitude boundary perturbations if $\hat k \gg \hat k^*$.

To better evaluate the effects of $\hat k$ on the possible evolutions of the reconnection process, we plot in Fig. \ref{fig6} the boundaries between the scenarios (1)-(3) in a parameter space map $(\hat \Psi_0, S)$ (as in Figs. \ref{fig2} and \ref{fig4}) for fixed $P_m = 5$ but different values of $\hat k$. The maximum area of existence of scenario (2) occurs for $\hat k \sim 1$, while for $\hat k \ll 1$ and $\hat k \gg 1$ the scenario (2) appears for a very limited range of ($\hat \Psi_0, \hat k$) parameters. Note also that the scenario (3) is greatly facilitated in the case of very large perturbation wave numbers ($\hat k \ll 1$) while the scenario (3) is facilitated by relatively large amplitude perturbations with ($\hat k \lesssim 1$).

\section{Discussion}\label{sec:discussion}

The introduction of a new type of phase/scenario diagrams that include explicitly the effects of the external drive has allowed us to graphically organize in a detailed way the possible evolutions of forced magnetic reconnection processes in collisional plasmas. In contrast to previous versions of the phase diagrams \citep[]{JD_2011, HBS_2011, DR_2012, HB_2013, CD_2013, KA_2013}, this new representation highlights regions of the parameter space ($\hat \Psi_0, \hat k$, $S$, $P_m$) in which reconnection is a slow diffusive process (Sec. \ref{secHK}) in addition to regions where reconnection can be fast (Secs. \ref{secWB} and \ref{our_scenario}). We recall that by fast we mean that the out-of-plane inductive electric field at the $X$-point is a significant fraction of the one evaluated upstream of the reconnection layer. 
We also emphasize that this kind of diagrams respond to the criticism moved by \citet[]{CD_2013} concerning the fact that these diagrams are not able of taking into account the dynamical evolution of the reconnection process from a slow to a fast regime inside of a given region of the parameter space. Indeed, scenarios (1)-(3) describe the forced magnetic reconnection process from the current sheet formation all the way to their specific nonlinear evolution.

We would like to remark that while the proposed parameter space diagrams represent a valid way to summarize the current knowledge of the forced magnetic reconnection dynamics in a collisional plasma, there are a number of conditions that may significantly affect the reconnection process which have not been addressed in this paper. 
For instance, two-fluid/kinetic effects should be considered if the length scale associated with the width of the reconnecting current sheet becomes of the order or smaller than the characteristic length scales of these effects. In fact, in antiparallel reconnection (i.e., in the absence of a guide magnetic field), Hall effects \citep[]{Birn_2001, SimaChac2008} are known to enhance the reconnection rate, as well as effects associated to finite electron inertia \citep[]{OttPor93, CA_2014}, electron pressure \citep[]{Kleva95, Grasso99} and ion gyration \citep[]{Comisso2013} are known to increase the reconnection rate in the presence of a strong guide field. 
We would like to remark also that a common condition in many physical systems is the presence of velocity flows, which are known to suppress the reconnection \citep[]{Fitz_NF1993,Waelb_NF2012} or to alter the reconnection rate \citep[]{Cassak_2011, Tassi_2014}. In this case our analysis should be extended by considering also the effects of a plasma flow on the reconnection dynamics. Similarly, also the effects of turbulence should be considered \citep[]{Servidio_2009,KA_2013} in order to obtain a more complete description of the magnetic reconnection dynamics.

Finally, it is important to recall that all the presented diagrams of magnetic reconnection are based on two-dimensional models and simulations. At present, the knowledge of how magnetic reconnection evolve in large three-dimensional systems is still far behind our understanding of what happens in two-dimensional systems. Therefore, despite the great progress achieved in recent years \citep[]{Borg_2005,Yin_2008,Daugh_2011,Wyper_2014}, other work is needed in this direction before we can implement a phase diagram description of three-dimensional magnetic reconnection.

\acknowledgments

The authors would like to acknowledge fruitful conversations with Richard Fitzpatrick and Enzo Lazzaro. 
This work was carried out under the Contract of Association Euratom-ENEA and was also supported by the U.S. Department of Energy under Contract No. DE-FG02-04ER-54742.




\end{document}